\documentclass[11pt]{article}
\pdfoutput=1

\usepackage{latexsym}
\usepackage{amsthm}
\usepackage{amsmath}
\usepackage{graphicx}
\usepackage{amssymb}
\usepackage{jcappub}
\usepackage[utf8]{inputenc}

\def\mpl{M_{\rm Pl}}
\newcommand{\be}{\begin{equation}}
\newcommand{\ee}{\end{equation}}
\newcommand{\bea}{\begin{eqnarray}}
\newcommand{\eea}{\end{eqnarray}}

\newcommand{\comment}[1]{}
\newcommand{\expect}[1]{\left\langle #1 \right\rangle}
\newcommand{\mbf}[1]{\mathbf #1}

\def\ep{\epsilon}

\def\simleq{\; \raise0.3ex\hbox{$<$\kern-0.75em
      \raise-1.1ex\hbox{$\sim$}}\; }
   \def\simgeq{\; \raise0.3ex\hbox{$>$\kern-0.75em
      \raise-1.1ex\hbox{$\sim$}}\; }

\begin{document}

\vspace*{0. cm}
\hfill CERN-PH-TH-2015-187
\vspace{20pt}

\begin{center}

{\LARGE\bf Effective Planck Mass and the Scale of Inflation}

\vskip 1cm

{\large \bf Matthew Kleban,$^a$
Mehrdad Mirbabayi,$^b$
and Massimo Porrati $^{a,c}$}

\vskip 0.5cm

{\large {$^a$\em Center for Cosmology and Particle Physics,  New York University}}

\vskip 0.2 cm
{\large {$^b$\em Institute for Advanced Study, Princeton}}

\vskip 0.2 cm
{\large {$^c$\em CERN PH-TH,
CH 1211, Geneva 23, Switzerland\footnote{Until September 1, 2015; on sabbatical leave from NYU.}}}
\end{center}

\vspace{.8cm}

\hrule \vspace{0.3cm}
{\noindent \textbf{Abstract} \\[0.3cm]
\noindent A recent paper argued that it is  not possible to infer the energy scale of inflation from the amplitude of tensor fluctuations in the Cosmic Microwave Background, because  the usual connection  is substantially altered if there are a large number of universally coupled fields present during inflation, with mass less than the inflationary Hubble scale.  We give a simple argument demonstrating that this is incorrect.
}
 \vspace{0.3cm}
\hrule

\section{Introduction} The energy scale of inflation remains very poorly constrained by observation.  The upper limit comes  from the fact that the amplitude of tensor modes in the Cosmic Microwave Background (CMB) is directly related to the scale of inflation.  In particular
$$
V^{1/4} \approx \left( {r \over 0.01}\right)^{1/4} \times 10^{16} {\rm \,  GeV},
$$
 where $V$ is the energy density at the time the  modes exited the horizon and $r$ is the ratio of tensor to scalar power.  Hence the current constraint on the amplitude of tensor fluctuations, $r \simleq 0.1$ \cite{Ade:2015lrj}, puts an upper limit on the scale of inflation.

 The constraint on $r$ is forecast to improve by  several orders of magnitude over the next few years, possibly down to a standard deviation $\sigma_r \sim 10^{-4}$ with next generation satellite experiments \cite{Creminelli:2015oda}.   If the simplest high-scale inflation models are correct, tensor modes will be discovered with high significance.  A lack of discovery would put a powerful constraint on the inflationary energy scale,  ruling out many models.

Clearly, the relationship between the scale of inflation and the amplitude of tensor modes is  of first importance.  There have been several attempts to demonstrate that the amplitude of tensor modes can be enhanced through various mechanisms involving extra degrees of freedom that are produced during inflation \cite{Senatore:2011sp, Cook:2011hg, Barnaby:2012xt}. This appears to be possible, although it requires a relatively complex mechanism and some tuning, and is often associated with large non-Gaussianity \cite{Ozsoy,Mirbabayi:2014jqa}. There is another source of degeneracy if the tensor modes do not propagate on the light-cone during inflation ($c_t\neq 1$). Again one would generically expect large non-Gaussianity since as shown in \cite{Creminelli} the cosmological predictions of any such model coincide with another model with $\tilde c_t=1$ and $\tilde c_s = c_s/c_t$, where $c_s$ is the propagation speed of scalar fluctuations and its deviation from unity is correlated with non-Gaussianity \cite{Cheung}.

Very recently, a  simple idea emerged:  that the presence of a large number of universally coupled spin 2 fields of mass less or of the order of the expansion rate $H_*$ could change the effective strength of gravitational interactions during inflation, thereby substantially affecting the connection between the amplitude of tensor modes and the scale of inflation \cite{Antoniadis}. More precisely, in the presence of $N$ such universally coupled fields (including the graviton) the authors of \cite{Antoniadis} conclude that tensor perturbations are given by
\be\label{M*}
\expect{\gamma^s_{\mbf k}\gamma^{s'}_{\mbf k'}}_{\rm vac}=(2\pi)^3\delta^{(3)}(\mbf k +\mbf k')\delta^{ss'} 
\frac{H^2_*}{M_*^2k^3},
\qquad M_*^2 \equiv \frac{\mpl^2}{N}.
\ee
If the mass of these fields satisifies TeV $\ll m \simleq H_* $, they would have no detectable effect on collider or gravitational experiments, but for large $N$ would have a strong influence during inflation.

\section{Counterarguments}

In this note we give two counterarguments and conclude that the connection between the tensor modes and $H_*$ cannot be changed in the presence of such fields, regardless of $N$.

The first counterargument is extremely simple, and invalidates the  assumption of having any light (universally coupled) spin-2 particles during inflation, let alone a large number. Light tensor modes are forbidden in de Sitter space by the Higuchi bound \cite{Higuchi}, which states that massive spin-2 particles of mass $m^2 <2 H_*^2$  have a ghost-like longitudinal excitation. The Higuchi bound is a consequence  of group theory: there is no unitary massive spin-2 representation of the $4$d de Sitter isometry group with $m^2<2 H_*^2$~\cite{dS1,dS2,dS3,Deser:2001us}. 
 The bound holds in any background with  $4$d $SO(1,4)$ isometries, regardless of its origin. In particular, 
 it puts a powerful, {\em universal} constraint on de Sitter compactifications of higher dimensional gravitational theories.  This is true regardless of any  ``warping" or other details (see \cite{Garriga} for one explicit example).  If one defines the ``size" of the extra dimensions by the inverse mass of the lightest Kaluza-Klein resonance of the graviton, the Higuchi bound forbids any compactification in which the extra dimensions are larger than and ${\cal O}(1)$ factor times the de Sitter radius.\footnote{M.K.~thanks N.~Arkani-Hamed for discussions on this point.}

The generalization to inflation, which deviates from exact de Sitter by the slow-roll parameter $\ep = -\dot H_*/H_*^2 \ll 1$, is by the assumption of continuity  of the spectrum  -- a reasonable assumption given that ghost instabilities are effectively instantaneous. Moreover, if the spectrum of spin 2 fields was gapped at $\epsilon=0$ and ungapped
at any $\epsilon\neq 0$, the whole scenario of slow-roll inflation would be put into question, since it is based  on the 
well-verified assumption that physics in a background with $\dot H_*/H_*^2 \ll 1$ is well approximated by physics on a de 
Sitter background.

Heavier spin-2 particles are dynamically less relevant at energy scales of order $H_*$, and for these the assumption that the effective gravitational strength at distances of order $H_*^{-1}$ increases proportionally to their multiplicity is no longer valid. 

One can still consider the case of having several universally coupled scalar fields which are much lighter than $H_*$, or assume that many massive spin-2 fields are accumulated just above $\sqrt{2} H_*$. It is then conceivable to have a stronger force between (non-relativistic) sources and an effectively larger gravitational strength $M_*^{-2}$. However, the following argument rules out any direct connection between $M_*$ and the cosmologically observable tensor fluctuations.


In order to discuss the scale of inflation one first needs to fix the system of units, and in particular the conformal frame for the metric. What we mean by the scale of inflation is its value in the Einstein frame, in which the Einstein-Hilbert action is normalized by the Planck mass. In this frame, an increase in the strength of attraction between two sources can arise from the exchange of new degrees of freedom (whose fluctuations about their VEV is collectively called $\chi_a$) in addition to the massless graviton $h_{\mu\nu}$. The dynamics of the latter is still governed to lowest order in derivatives by the Einstein-Hilbert action:
\be
S_{\rm EH} =\frac{1}{2}\mpl^2 \int d^4x \sqrt{-g} R.
\ee
This action fully fixes the vacuum fluctuations of $h_{\mu\nu}$ in a quasi-de Sitter background; the transverse-traceless fluctuations are given by \eqref{M*} but with $M_*\to \mpl$. Higher order curvature corrections are suppressed by factors of $H_*^2/M_*^2$, and hence they are negligible as long as gravity is weakly coupled.
Specifically, a term $\propto R^2$
 can be removed by a well-known field redefinition~\cite{whitt} while a term $cR_{\mu\nu}R^{\mu\nu}$ sets the 
cutoff of the theory at $M_{Pl}/\sqrt{c}$. Corrections $\delta P_T  $ to the tree-level power 
spectrum of tensor modes, $P_T$, are at best 
$O(cH^2/M_{Pl}^2)P_T$. Clearly  $\delta P_T \ll P_T$ for the curvature expansion to make sense.

Next we discuss the fluctuations of other light degrees of freedom $\chi_a$ during inflation and argue that they have no observable effect on the tensor spectrum. For massive spin-2 fields satisfying the Higuchi bound the super-horizon fluctuations decay during inflation according to
\be\label{decay}
\chi(t) \propto \left\{\begin{array}{ll} a^{-3/2+\sqrt{9/4-m_\chi^2/H_*^2}}& \mbox{if $\sqrt{2} H_* < m_\chi < 3 H_*/2$};\\
a^{-3/2} &\mbox{if $m_\chi \geq 3 H_*/2$,}\end{array}\right.
\ee
where $a$ is the scale factor.   This rapid decay means that the fluctuations in the massive tensors do not produce observable features in the CMB.

On the other hand for minimally coupled scalar fields the fluctuations do not decay significantly during inflation if $m_\chi \ll H_*$. However, if these fields have universal coupling to matter with gravitational strength, astrophysical observations and laboratory experiments constrain them to be much heavier than the Hubble scale at recombination. The super-horizon fluctuations of such fields start to decay in a  fashion similar to \eqref{decay} once the expansion rate falls below their mass during the cosmic evolution. This behavior falls in the category of multifield inflationary models, where isocurvature fluctuations of a scalar field $\chi$ can convert into adiabatic fluctuations and enhance the scalar power, but have no effect on the tensor fluctuations.  In any case this can only reduce the tensor to scalar ratio $r$, and does not alter the connection between $H_*$ and the tensor power.  

Vectors or higher spin fields also have no effect, because vectors decay outside the horizon too rapidly to be of any relevance, and  fields with spin higher than 2 and mass less than $\sim H_*$ lead to ghost instabilities similar to those of spin 2 \cite{Deser:2001us}.

In conclusion, contrary to the claims of \cite{Antoniadis}, $N \gg 1$ universally coupled fields cannot alter the connection between the amplitude of tensor fluctuations and the scale of inflation.

\section*{Acknowledgements}
It is a pleasure to thank Nima Arkani-Hamed  and Paolo Creminelli for useful discussions.
The work of MK is supported in part by the NSF  through grant PHY-1214302. MM acknowledges support from NSF Grants 
PHY-1314311, PHY-0855425, and PHY-1066293, and the hospitality of the Aspen Center for Physics. The work 
of MP is supported in part by NSF grant PHY-1316452.

\bibliography{aabib}

\providecommand{\href}[2]{#2}\begingroup\raggedright\begin{thebibliography}{10}

\bibitem{Ade:2015lrj}
{\bfseries Planck} Collaboration, {\sc P.~A.~R. Ade} et~al., ``{Planck 2015
  results. XX. Constraints on inflation},''
\href{http://arxiv.org/abs/1502.02114}{{\ttfamily arXiv:1502.02114
  [astro-ph.CO]}}.

\bibitem{Creminelli:2015oda}
{\sc P.~Creminelli}, {\sc D.~L. Nacir}, {\sc M.~Simonovi\'c}, {\sc
  G.~Trevisan}, and {\sc M.~Zaldarriaga}, ``{Detecting Primordial $B$-Modes
  after Planck},''
\href{http://arxiv.org/abs/1502.01983}{{\ttfamily arXiv:1502.01983
  [astro-ph.CO]}}.

\bibitem{Senatore:2011sp}
{\sc L.~Senatore}, {\sc E.~Silverstein}, and {\sc M.~Zaldarriaga}, ``{New
  Sources of Gravitational Waves during Inflation},''
  \href{http://dx.doi.org/10.1088/1475-7516/2014/08/016}{{\em JCAP} {\bfseries
  1408} (2014) 016},
\href{http://arxiv.org/abs/1109.0542}{{\ttfamily arXiv:1109.0542 [hep-th]}}.

\bibitem{Cook:2011hg}
{\sc J.~L. Cook} and {\sc L.~Sorbo}, ``{Particle production during inflation
  and gravitational waves detectable by ground-based interferometers},''
  \href{http://dx.doi.org/10.1103/PhysRevD.86.069901,
  10.1103/PhysRevD.85.023534}{{\em Phys. Rev.} {\bfseries D85} (2012) 023534},
  \href{http://arxiv.org/abs/1109.0022}{{\ttfamily arXiv:1109.0022
  [astro-ph.CO]}}.
[Erratum: Phys. Rev.D86,069901(2012)].

\bibitem{Barnaby:2012xt}
{\sc N.~Barnaby}, {\sc J.~Moxon}, {\sc R.~Namba}, {\sc M.~Peloso}, {\sc
  G.~Shiu}, and {\sc P.~Zhou}, ``{Gravity waves and non-Gaussian features from
  particle production in a sector gravitationally coupled to the inflaton},''
  \href{http://dx.doi.org/10.1103/PhysRevD.86.103508}{{\em Phys. Rev.}
  {\bfseries D86} (2012) 103508},
\href{http://arxiv.org/abs/1206.6117}{{\ttfamily arXiv:1206.6117
  [astro-ph.CO]}}.

\bibitem{Ozsoy}
{\sc O.~{\"O}zsoy}, {\sc K.~Sinha}, and {\sc S.~Watson}, ``{How Well Can We
  Really Determine the Scale of Inflation?},''
  \href{http://dx.doi.org/10.1103/PhysRevD.91.103509}{{\em Phys. Rev.}
  {\bfseries D91} no.~10, (2015) 103509},
\href{http://arxiv.org/abs/1410.0016}{{\ttfamily arXiv:1410.0016 [hep-th]}}.

\bibitem{Mirbabayi:2014jqa}
{\sc M.~Mirbabayi}, {\sc L.~Senatore}, {\sc E.~Silverstein}, and {\sc
  M.~Zaldarriaga}, ``{Gravitational Waves and the Scale of Inflation},''
  \href{http://dx.doi.org/10.1103/PhysRevD.91.063518}{{\em Phys. Rev.}
  {\bfseries D91} (2015) 063518},
\href{http://arxiv.org/abs/1412.0665}{{\ttfamily arXiv:1412.0665 [hep-th]}}.

\bibitem{Creminelli}
{\sc P.~Creminelli}, {\sc J.~Gleyzes}, {\sc J.~Noreña}, and {\sc F.~Vernizzi},
  ``{Resilience of the standard predictions for primordial tensor modes},''
  \href{http://dx.doi.org/10.1103/PhysRevLett.113.231301}{{\em Phys. Rev.
  Lett.} {\bfseries 113} no.~23, (2014) 231301},
\href{http://arxiv.org/abs/1407.8439}{{\ttfamily arXiv:1407.8439
  [astro-ph.CO]}}.

\bibitem{Cheung}
{\sc C.~Cheung}, {\sc P.~Creminelli}, {\sc A.~L. Fitzpatrick}, {\sc J.~Kaplan},
  and {\sc L.~Senatore}, ``{The Effective Field Theory of Inflation},''
  \href{http://dx.doi.org/10.1088/1126-6708/2008/03/014}{{\em JHEP} {\bfseries
  03} (2008) 014},
\href{http://arxiv.org/abs/0709.0293}{{\ttfamily arXiv:0709.0293 [hep-th]}}.

\bibitem{Antoniadis}
{\sc I.~Antoniadis} and {\sc S.~P. Patil}, ``{The Effective Planck Mass and the
  Scale of Inflation},''
  \href{http://dx.doi.org/10.1140/epjc/s10052-015-3411-z}{{\em Eur. Phys. J.}
  {\bfseries C75} (2015) 182},
\href{http://arxiv.org/abs/1410.8845}{{\ttfamily arXiv:1410.8845 [hep-th]}}.

\bibitem{Higuchi}
{\sc A.~Higuchi}, ``{Forbidden Mass Range for Spin-2 Field Theory in De Sitter
  Space-time},''
\href{http://dx.doi.org/10.1016/0550-3213(87)90691-2}{{\em Nucl. Phys.}
  {\bfseries B282} (1987) 397}.

\bibitem{dS1}
{\sc L.~Thomas}, ``{On unitary representations of the group of De Sitter
  space},'' {\em Annals of Math.} {\bfseries 42} (1941) 113.

\bibitem{dS2}
{\sc T.~Newton}, ``{A note on the representations of the De Sitter group},''
  {\em Annals of Math.} {\bfseries 51} (1950) 730.

\bibitem{dS3}
{\sc J.~Dixmier}, ``{Repr\'esentations int\'egrables du group de De Sitter},''
  {\em Bulletin S.M.F.} {\bfseries 89} (1961) 9.

\bibitem{Deser:2001us}
{\sc S.~Deser} and {\sc A.~Waldron}, ``{Partial masslessness of higher spins in
  (A)dS},'' \href{http://dx.doi.org/10.1016/S0550-3213(01)00212-7}{{\em Nucl.
  Phys.} {\bfseries B607} (2001) 577--604},
\href{http://arxiv.org/abs/hep-th/0103198}{{\ttfamily arXiv:hep-th/0103198
  [hep-th]}}.

\bibitem{Garriga}
{\sc J.~Garriga} and {\sc M.~Sasaki}, ``{Brane world creation and black
  holes},'' \href{http://dx.doi.org/10.1103/PhysRevD.62.043523}{{\em Phys.
  Rev.} {\bfseries D62} (2000) 043523},
\href{http://arxiv.org/abs/hep-th/9912118}{{\ttfamily arXiv:hep-th/9912118
  [hep-th]}}.

\bibitem{whitt}
{\sc B.~Whitt}, ``{Fourth Order Gravity as General Relativity Plus Matter},''
  {\em Phys. Lett.} {\bfseries B145} (1984) 176.

\end{thebibliography}\endgroup

\vskip .5cm
\vskip .5 cm

\end{document}